\documentclass [12pt] {article}
\usepackage{graphicx}
\usepackage{epsfig}
\def\lsim{\raise0.3ex\hbox{$<$\kern-0.75em\raise-1.1ex\hbox{$\sim$}}}
\def\gsim{\raise0.3ex\hbox{$>$\kern-0.75em\raise-1.1ex\hbox{$\sim$}}}
\begin {document}

\vskip -0.1cm
\begin{center}
{\Large {\bf  STRING JUNCTION EFFECTS FOR FORWARD \newline \vskip 0.1cm
AND CENTRAL BARYON PRODUCTION \newline \vskip -0.1cm
IN HADRON-NUCLEUS COLLISIONS}} \\

\vskip 1.5 truecm
{\bf F. Bopp and Yu. M. Shabelski$^1$}\\
\vskip 0.5 truecm
Siegen University, Germany \\
E-mail: Bopp@physik.uni-siegen.de
\end{center}
\vskip 2. truecm
\begin{center}
{\bf ABSTRACT}
\end{center}

The process of baryon number transfer due to string junction propagation
in rapidity space is considered. It leads to a significant effect in the
net baryon production in $pA$ collisions at mid-rapidities and an even
more significant effect in the forward hemisphere 
for  the cases of $\pi A$
interactions. The results of numerical calculations in the framework of
the Quark-Gluon String Model are in reasonable agreement with the data.
Special consideration is given to  $\Lambda$ produced 
in $\pi^- A$ collisions extracted from data of WA89 Collaboration.

\vskip 1cm


\vskip 2.5 truecm

\noindent {\small$^1$Permanent address: Petersburg Nuclear Physics Institute,
Gatchina, St.Petersburg, Russia \newline \vspace{1em}
E-mail: shabelsk@thd.pnpi.spb.ru}

\newpage
\pagestyle{plain}
\noindent{\bf 1. Introduction}
\vskip 0.4 truecm

The Quark--Gluon String Model (QGSM) and the Dual Parton Model (DPM) are
based on the Dual Topological Unitarization (DTU) and 
describe many features of high energy production processes, both in
hadron--nucleon and hadron--nucleus collisions \cite{KTM}-\cite{Sh1} 
quite reasonably. 
In these models high energy interactions proceed via the exchange
of one or several Pomerons and all elastic and inelastic processes
result from cutting through or between Pomerons \cite{AGK} using the
reggeon counting rules \cite{Kai}. 
Each cut Pomeron leads to the production of two strings  of secondaries. 
Inclusive
spectra of hadrons are 
in this way related to the corresponding fragmentation
functions of quarks and diquarks at the end of the strings.

In the string models baryons are considered as configurations consisting
of three strings attached to three valence quarks and connected in a point
called ``string junction" (SJ) \cite{Artru}-\cite{MRV}. 
String junctions are modelled as objects of nonperturbative QCD.
It is very interesting to understand their role in the dynamics
of high-energy hadronic interactions, in particular in the processes of
baryon number transfer \cite{IOT1}-\cite{Bopp}. 

Important results connected with the transfer
of baryon charge over long rapidity distances 
in string models 
were obtained in \cite{ACKS} and 
following papers \cite{SJ1,Olga}. 
In the present paper we consider such  processes with transfers 
away from nuclear targets. 

In nuclear targets these discussed transfers are enhanced due
to two reasons. First, the usual production of secondaries (which can
be considered as a background for string junction effects) in the beam
fragmentation region is suppressed due to nuclear absorption
\cite{KTMS,Sh2,CK,ASS}. Second, the probability of the baryon number
transfer should be proportional to the number of inelastic interactions
in the nuclear matter, $\langle \nu \rangle_{hA}$.  In the case of baryon
beams the SJ effects are the most visible in the central (midrapidity)
region \cite{ACKS,SJ1}. Most interesting for
meson--nucleus collisions is the forward region.  
The rapidity range available for SJ effects is drastically expanded in this
way.

\vskip 0.4 truecm
\noindent{\bf 2. Baryon as 3q + SJ system}
\vskip 0.4 truecm

In QCD hadrons are composite bound state configurations built up from the
quark $\psi_i(x), i = 1,...N_c$ and gluon $G^{\mu}_a(x), a = 1,...,N_c^2-1$
fields. In the string models baryons are considered as configurations
consisting of three strings attached to three valence quarks and connected in
a point called the ``string junction" (SJ) \cite{Artru}-\cite{MRV}. The 
correspondent wave function can be written as
\begin{equation}
B = \psi_i(x_1) \psi_j(x_2)\psi_k(x_3) J^{ijk} \;,
\end{equation}

\begin{equation}
J^{ijk} = \Phi^i_{i'}(x_1,x)\Phi_{j'}^j(x_2,x)\Phi^k_{k'}(x_3,x)
\epsilon^{i'j'k'} \;,
\end{equation}

\begin{equation}
\Phi^i_{i'}(x_1,x) = \left[ T\exp \left(g \int_{P(x_1,x)} A_{\mu}(z)
dz^{\mu}\right) \right]^i_{i'} \;.
\end{equation}
In the last equation $P(x_1,x)$ represents a path from $x_1$ to $x$
which looks like an open string with ends in $x_1$ and $x$ (see Fig.~1b).
Such baryon wave function can be defined as a ``star" or ``Y" shape
which is preferable \cite{Artru,RV} in comparison with ``triangle" 
(``ring") or ``\,$\Delta$" shape. 
The meson wave function 
\begin{equation}
M = \psi_i(x_1) \Phi^i_{i'}(x_1,x_2)\psi^{i'}(x_2) \;.
\end{equation}
has the form of an ``open string" 
with quark resp. antiquark on its end. Quarkless states (glueballs, see Fig.~1c)
are quite natural in this approach.

\begin{figure}[htb]
\centering
\includegraphics[width=.5\hsize]{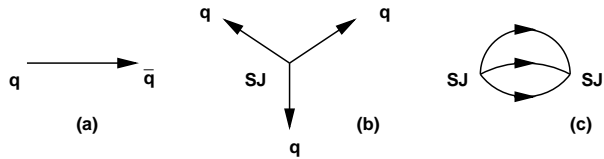}
\caption{Composite structure of a meson (a), a baryon (b) 
and a quarkless baryonium (c) in string models.
}
\end{figure}

Such a picture leads to several phenomenological predictions. 
In the additive quark model the dynamic of the strings is attributed to 
its ends. This means that 
a meson consists 
of two constituent quarks as depicted in Fig.~1a, and that   
a baryon consists 
of four constituent objects, three constituent
quarks and a junction line (SJ), as it is shown in Fig. 1b. 
The picture has an immediate consequence for the cross sections.
The ratio of nucleon--nucleon and meson--nucleon total cross sections
at high energies increases \cite{KKKY} in comparison with
classical result \cite{LF} of
$\sigma (N-N) / \sigma (\pi-N) = 3/2$. 
Accounting for the possibility of SJ interaction with a target one gets
\begin{equation}
\frac{\sigma (N-N)}{\sigma (\pi-N)} = \frac32 +
\frac{\sigma (SJ-N)}{2\sigma(q-N)} \;,
\end{equation}
where the additional term $\sigma (SJ-N) / (2\sigma(q-N))$ can be
estimated \cite{KKKY} to be equal $1/5 \pm  1/7$. This correction
results in better agreement \cite{AKNS} with experimental data.

Flavor dependent quark antiquark annihilation will introduce 
corrections to this simple picture.
The SJ annihilation in $B\bar{B}$ cross section, $\sigma_{ann}$, is not
necessarily the dominant effect. 
It is not equal to the difference 
$\Delta \sigma = \sigma^{tot}(B\bar{B}) - \sigma^{tot}(BB)$
\cite{RV}. 

The existence of SJ in a baryon structure changes the quark counting 
rules for reactions with large momenta transfer \cite{IOT1,Noda}. 

The reaction $\bar{p} p \to \bar{\Omega} \Omega$ can occur now without
breaking the OZI rules. The ratio of $\bar{\Omega}/\Omega$ production
for the collisions of non-strange hadrons is predicted to be smaller
than unity \cite{Noda} contrary to many models for multiparticle
production. This prediction is in agreement with the experimental data
\cite{ait} and their description in \cite{ACKS,SJ1}.
In the case of inclusive reactions the baryon number transfer to large
rapidity distances in hadron--nucleon reactions can be explained by SJ
diffusion \cite{ACKS}. 

In the paper we consider the effects of string junction
for hadron--nucleus inelastic interactions.


\vskip 0.4 truecm
\noindent{\bf 3. Production of secondaries on nuclear targets in QGSM}
\vskip 0.4 truecm

As mentioned above, high energy hadron--nucleon and hadron--nucleus
interactions are considered in the QGSM and in DPM as proceeding via
the exchange of one or several Pomerons. 
As said, each Pomeron corresponds
to a cylindrical diagram, and thus, when cutting a Pomeron two showers
of secondaries are produced. The inclusive spectrum of secondaries is
determined by the convolution of diquark, valence quark and sea quark
distributions $u(x,n)$ in the incident particles and the fragmentation
functions $G(z)$ of quarks and diquarks into secondary hadrons.

The diquark and quark distribution functions depend on the number $n$ of cut
Pomerons in the considered diagram. In the following we use the formalism of
QGSM.

In the case of a nucleon target the inclusive spectrum of a secondary
hadron $h$ has the form \cite{KTM}:

\begin{equation} \frac{x_E}{\sigma_{inel}}
\frac{d\sigma}{dx} =\sum_{n=1}^{\infty}w_{n}\phi_{n}^{h}(x)\ \ ,
\end{equation}
where $x=2p_{\|}/\sqrt{s}$ is the Feynman variable $x_F$ 
and where $x_E=2E/\sqrt{s}$

The functions $\phi_{n}^{h}(x)$ determine the contribution of
diagrams with $n$ cut Pomerons and $w_{n}$ is the probability of this
process. Here we neglect the contributions of diffraction dissociation
processes which are comparatively small in most of the processes
considered below. It can be accounted for separately \cite{KTM,2r}.

In this paper we consider mainly the  $\pi A$ interactions, so we
present the formulae for 
$\pi p$ collisions
\begin{equation}
\phi_{\pi p}^{h}(x) = f_{\bar{q}}^{h}(x_{+},n)f_{q}^{h}(x_{-},n) +
f_{q}^{h}(x_{+},n)f_{qq}^{h}(x_{-},n) +
2(n-1)f_{s}^{h}(x_{+},n)f_{s}^{h}(x_{-},n)\ \  ,
\end{equation}
\noindent where
\begin{equation}
x_{\pm} = \frac{1}{2}[\sqrt{4m_{T}^{2}/s+x^{2}}\pm{x}]\ \ ,
\end{equation}
where the transverse mass of the produced hadron is
$$m_T = \sqrt{m^2 + p^2_T}$$
and where $f_{qq}$, $f_{q}$ and $f_{s}$ correspond to the contributions of
diquarks, valence quarks and sea quarks, respectively. They are
determined by the convolution of the diquark and quark distributions
with the fragmentation functions, e.g.,
\begin{equation}
f_{qq}^{h}(x_{-},n) = \int_{x_{-}}^{1} u_{qq}(x_{1},n)\cdot
G_{qq}^{h}(x_{-}/x_{1}) \ dx_{1} \\ .
\end{equation}

In the case of nuclear targets we must consider the possibility of
one or several Pomeron cuts in each of the $\nu$ blobs of
hadron--nucleon inelastic interactions as well as cuts between Pomerons.
For example, for a $\pi A$ collision one of the cut Pomerons links a
valence antiquark and a valence quark of the projectile pion with a valence
quark and diquark of one target nucleon. 
The  other Pomerons link the
sea quark--antiquark pairs of the projectile pion with diquarks and
valence quarks of  other target nucleons or with sea quark--antiquark
pairs  of the target.

For example, one of the diagram for inelastic interaction with two
target nucleons is shown in Fig.~2. In the blob of the $\pi N_1$
inelastic interaction one Pomeron is cut, and in the blob of $\pi
N_2$ interaction two Pomerons are cut. It is essential to take
into account every possible Pomeron configuration and permutation
on all diagrams. The process shown in Fig.~2 satisfies the
condition that the absorptive parts of hadron--nucleus amplitude
are determined by the combinations of the absorptive parts of
hadron-nucleon interactions.

\vspace*{0.8cm}

\begin{figure}[htb]
\centering
\includegraphics[width=.5\hsize]{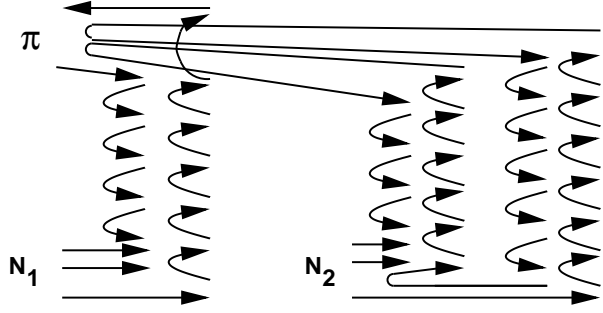}
\caption{One of the diagrams for inelastic interaction of an
incident pion with two target nucleons $N_1$ and $N_2$ in a $\pi
A$ collission. }
\end{figure}

In the case of inelastic interactions with $\nu$ target nucleons $n$ be
the total number of cut Pomerons in $hA$ collisions ($n \geq \nu$) and
let $n_i$ be the number of cut Pomerons connecting with the $i$-th
target nucleon ($1 \leq n_i \leq n-\nu+1$). We define the relative
weight of the contribution with $n_i$ cut Pomerons in every $hN$ blob
as $w^{hN}_{n_i}$. For the inclusive
spectrum of the secondary hadron $h$ produced in a $\pi A$
collision we obtain \cite{KTMS}
\begin{eqnarray}
\frac{x_E}{\sigma^{prod}_{\pi A}} \frac{d \sigma}{dx_F} & = &
\sum^A_{\nu=1} V^{(\nu)}_{\pi A} \left\{ \sum^{\infty}_{n=\nu}
\sum^{n-\nu+1}_{n_1 = 1} \cdot \cdot \cdot
\sum^{n-\nu+1}_{n_{\nu}=1} \prod^{\nu}_{l=1} w^{\pi N}_{n_l} \right.
\times \\ \nonumber
& \times & [f^h_{\bar{q}}(x_+,n)f^h_q(x_-,n_l) +
f^h_q(x_+,n)f^h_{qq}(x_-,n_l) + \\ \nonumber
& + & \sum^{2n_l-2}_{m=1} f^h_s(x_+,n)f^h_s(x_-,n_m)]
\left. \right\} \;,
\end{eqnarray}
where $V^{(\nu)}_{pA}$ is the probability of ``pure inelastic"
(non diffractive) interactions with $\nu$ target nucleons, and we should
account for all possible Pomeron permutation and the difference in quark
content of the protons and neutrons in the target.

In particular, the contribution of the diagram in Fig.~2 to the
inclusive spectrum is
\begin{eqnarray}
\frac{x_E}{\sigma^{prod}_{\pi A}} \frac{d \sigma}{dx_F} & = &
2 V^{(2)}_{\pi A} w^{\pi N_1}_1w^{\pi N_2}_2
\left\{ f^h_{\bar q}(x_+,3)f^h_q(x_-,1)\right. + \\ \nonumber
& + & f^h_q(x_+,3)f^h_{qq}(x_-,1) + f^h_s(x_+,n) [f^h_{qq}(x_-,2) +
f^h_q(x_-,2) + \\ \nonumber
& + & 2f^h_s(x_-,2)] \left. \right\} \;.
\end{eqnarray}

In the case of a nucleon beam the valence antiquark contributions of
incident particle should be changed by the contribution of valence
diquarks.

The diquark and quark distributions as well as the fragmentation
functions are determined from Regge intercepts. Their expressions are
given in Appendix 1 of \cite{ACKS} (see also \cite{SJ1}).

According to \cite{ACKS,SJ1} we account three possibilities that the
secondary baryon can consist of the SJ together with two valence and
one sea quarks (a), with one valence and two sea quarks (b) or with
three sea quarks (c). They are shown in Fig.~3.

\vspace*{-4cm}
\begin{figure}[htb]
\centering
\includegraphics[width=.55\hsize]{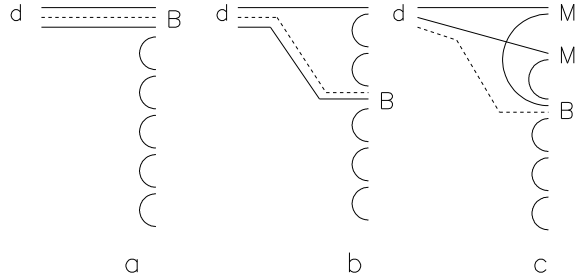}
\caption{QGSM diagrams describing secondary baryon $B$ production by
diquark $d$: initial SJ together with two valence quarks and one sea
quark (a), together with one valence quark and two sea quarks (b)
and together with three sea quarks (c).
}
\end{figure}

The fraction of the incident baryon energy carried by the secondary
baryon decreases from (a) to (c), whereas the mean rapidity gap between
the incident and secondary baryon increases.

The processes shown in Fig.~3a and 3b are the standard ones in QGSM
and DPM. They determine the main contribution to the multiplicity of
secondary baryons $B$ in the fragmentation region. The most interesting
for us at high energies is the case shown in  Fig.~3c, which leads to
the difference in baryon and antibaryon production at rapidities far
from the incident baryon (baryon charge diffusion in rapidity space).
In the case of secondary baryon production with mass $m$ in high energy
meson--nucleus collisions for positive (forward hemisphere) and not very
small $x_F$ the value $x_- \approx m_T^2 /(x_F s)$ and the correspondent
contribution to the target diquark fragmentation function (see Eq.~(9))
has the form
\begin{equation}
G_{SJ}^{h}(z) \sim \varepsilon z^{1 - \alpha_{SJ}} \;.
\end{equation}
Here $\varepsilon$ is the relative suppression of the discussed
contribution in comparison with the processes (a) and (b), and
$\alpha_{SJ}$ is the intercept of SJ Regge trajectory.
In the present calculations we use the values
$\varepsilon = 0.024,\  \alpha_{SJ} = 0.9,$and $ a_N = 1.33$ as in \cite{SJ1}.
Following to the modern experimental data, we increase the
portion of strange quarks in the sea, $S/L$ \cite{CS,ACKS} from
$S/L = 0.2$ to $S/L = 0.32$.

To illustrate the expected effects of SJ contributions we present in
Fig.~4 the predicted inclusive cross sections of
$\pi^-Cu \to \Lambda X$ and $\pi^-Cu \to \Omega^-X$ reactions with
(solid curves) and without (dashed curves) SJ contributions of 
Fig.~3c. 
These reactions were selected as the secondary
baryons and the correspondent antibaryons $\bar{\Lambda}$ and
$\overline{\Omega^+}$ have symmetrical quark states in respect to the
incident $\pi^-$. 
So the SJ contribution  equals to the difference
between solid and dashed curves in Fig.~4. It can be measured experimentally,
as the difference in $\Lambda - \bar{\Lambda}$, or
$\Omega^- - \overline{\Omega^+}$ production at high energies.

In general the SJ contribution shown in Fig.~3c increases the inclusive
cross sections of $\Lambda$ and $\Omega^-$ production. The spectra of
antibaryons are not affected. However, numerically these effects are 
rather small, for example, the mean multiplicity of secondary $\Lambda$ in
forward hemisphere should increase about 15\% which should be compensated
by the correspondent decrease of secondary nucleon multiplicity
in the target region.

\begin{figure}[htb]
\centering
\includegraphics[width=.5\hsize]{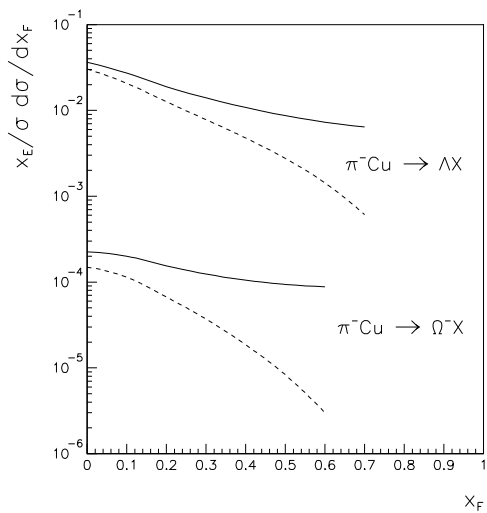}
\vskip -0.5cm
\caption{The QGSM predictions for the inclusive cross sections of
$\Lambda$ and $\Omega^-$ production in $\pi^-Cu$ collisions
at 400 GeV/c with (solid curves) and without (dashed curves) SJ
contributions.
}
\end{figure}


\vskip 0.9 truecm
\noindent{\bf 4. Comparison with the data}
\vskip 0.5 truecm

In Fig.~5 we show the data \cite{WA97} on the midrapidity inclusive
densities, $dn/dy$ at $\vert y_{c.m.} \vert < 0.5$ of secondary
$\Lambda$, $\bar{\Lambda}$, $\Xi^-$, $\overline{\Xi^+}$ and the sum
$\Omega^- + \overline{\Omega^+}$ produced in $pBe$ and $pPb$ collisions
at 158 GeV/c. These data are in reasonable agreement with the QGSM
calculations and this agreement is better with accounting the SJ
contributions for secondary baryons.

\begin{figure}[htb]
\centering
\includegraphics[width=.5\hsize]{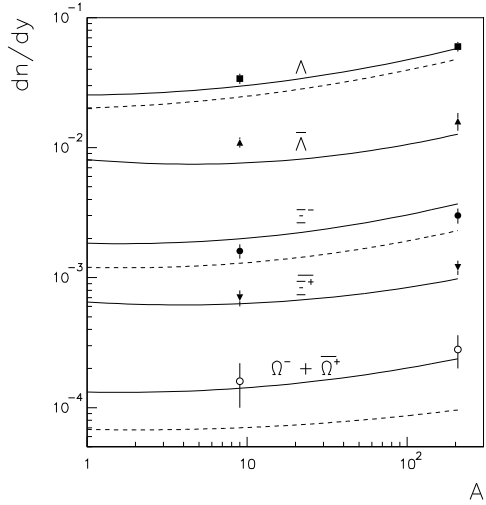}
\vskip -1.0cm
\caption{Yields of $\Lambda$ (closed squares), $\bar{\Lambda}$
(triangles), $\Xi^-$ (points), $\overline{\Xi^+}$ (turned over
triangles) and the sum $\Omega^- + \overline{\Omega^+}$ (stars)
per unit of rapidity at central rapidity as a function of the target
atomic weight for $pA$ collisions at 158 GeV/c. QGSM predictions with
SJ contribution are shown by solid curves and without SJ contribution by
dashed curves
.}
\end{figure}


In the case of pion--nucleus collisions at fixed target energies the SJ
effects are more important. In Fig.~6 we present the NA49 Coll. data
\cite{NA49} on the $x_F$ distributions of net protons $(p - \bar{p})$
produced in $\pi p$ and $\pi Pb$ interactions at $\sqrt{s}$ = 17.2 GeV.
The beam $\pi $ is determined in \cite{NA49} as $(\pi^+ + \pi^-)/2$.
The data are described rather good with the full model accounting the SJ diffusion
(solid curves in Fig.~6) and the variant without SJ contribution
(dashed curves) underestimates the data several times at $x_F > 0.1$.

\begin{figure}[htb]
\centering
\vskip -0.5cm
\includegraphics[width=.55\hsize]{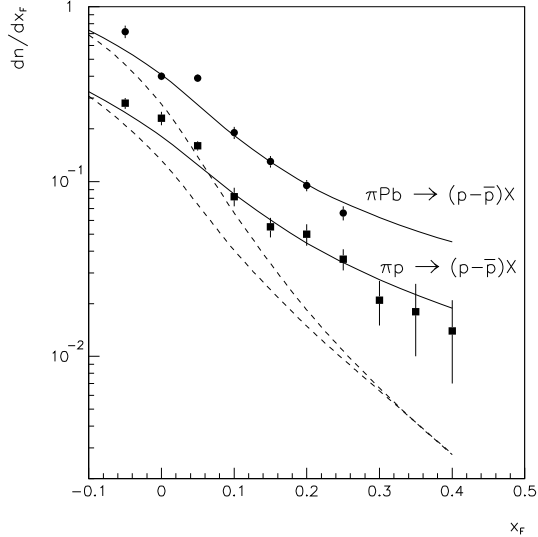}
\vskip -1.0cm
\caption{Feynman-$x$ distributions of net protons $(p - \bar{p})$
produced in $\pi p$ (squares) and $\pi Pb$ (points) interactions at
$\sqrt{s}$ = 17.2 GeV. Solid and dashed curves show the QGSM description
with and without SJ contribution, respectively
.}
\end{figure}

The experimental data of WA89 Coll. \cite{WA89} on $\Lambda$, $\Xi^-$
$\bar{\Lambda}$ and $\overline{\Xi^+}$ production from $C$ and $Cu$ targets
by 345 GeV/c $\pi^-$ beam are shown in Fig.~7. The yields of
secondary hyperons are in agreement with QGSM predictions accounting
for the SJ contributions (solid curves in Fig.~7a, b). The calculations
without SJ contributions (dashed curves) are in disagreement with
the data.

\begin{figure}[h]
\centering
\vskip -2.0cm
\includegraphics[width=.48\hsize]{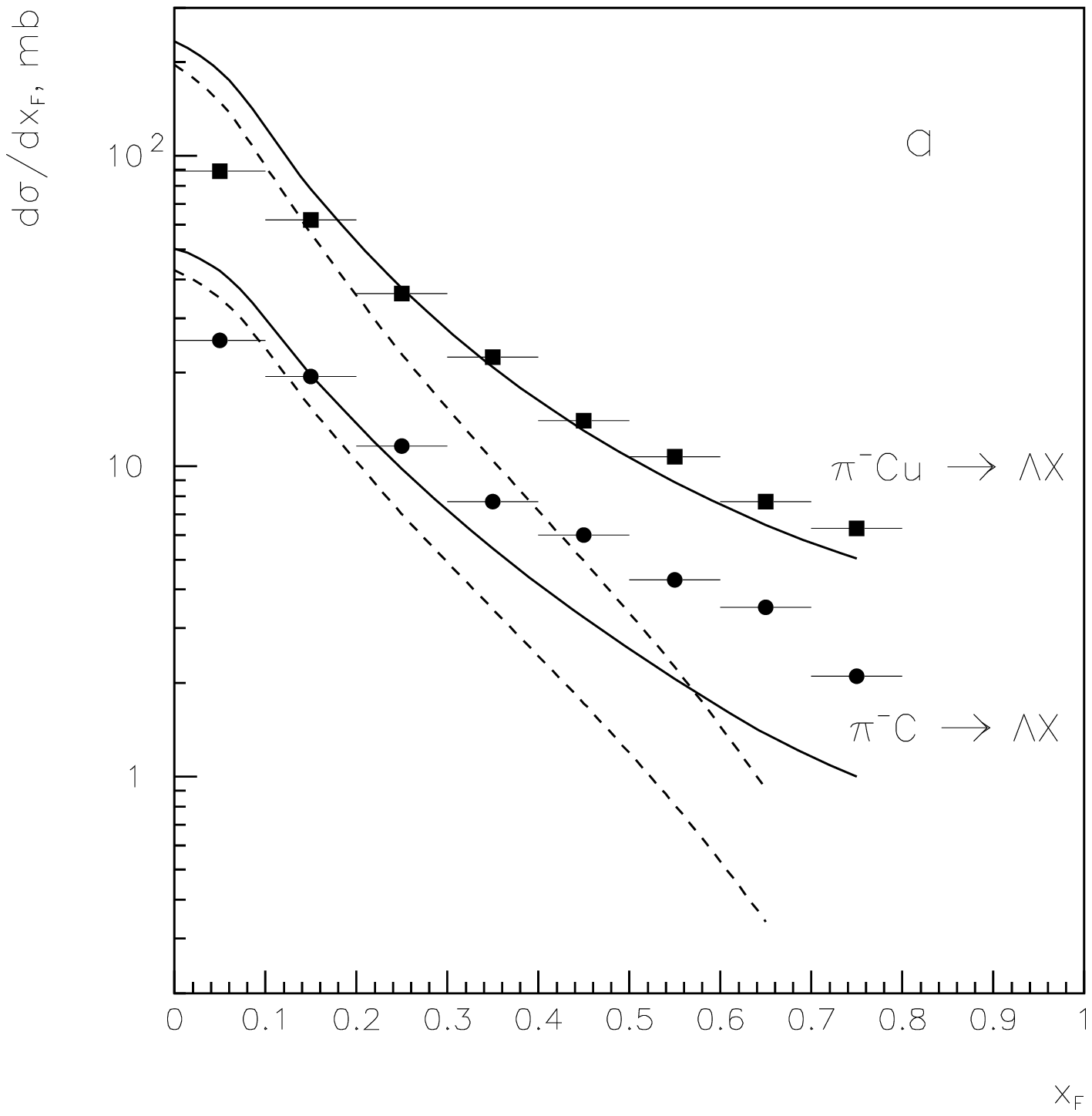}
\includegraphics[width=.48\hsize]{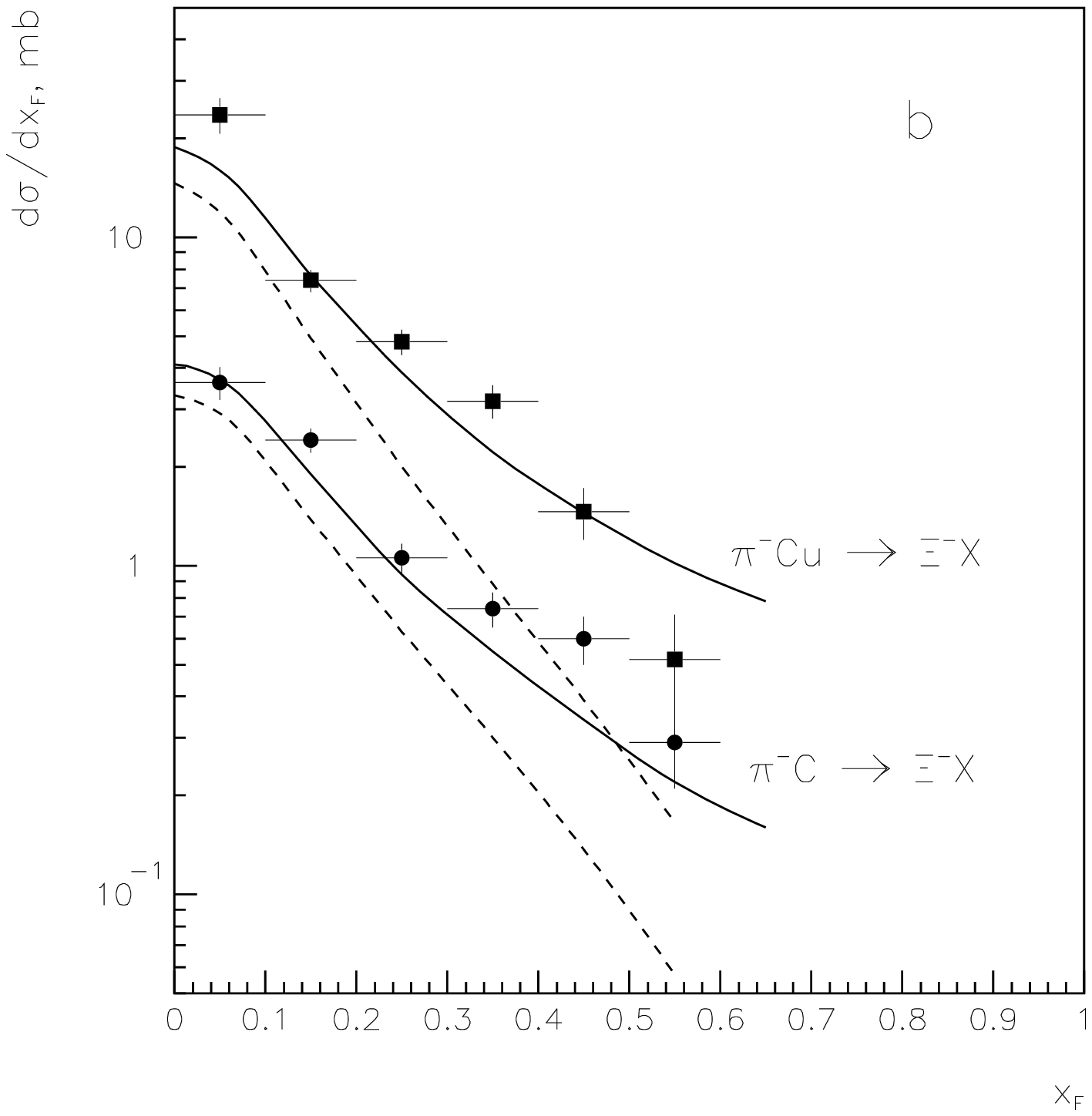}
\vskip -1cm
\includegraphics[width=.48\hsize]{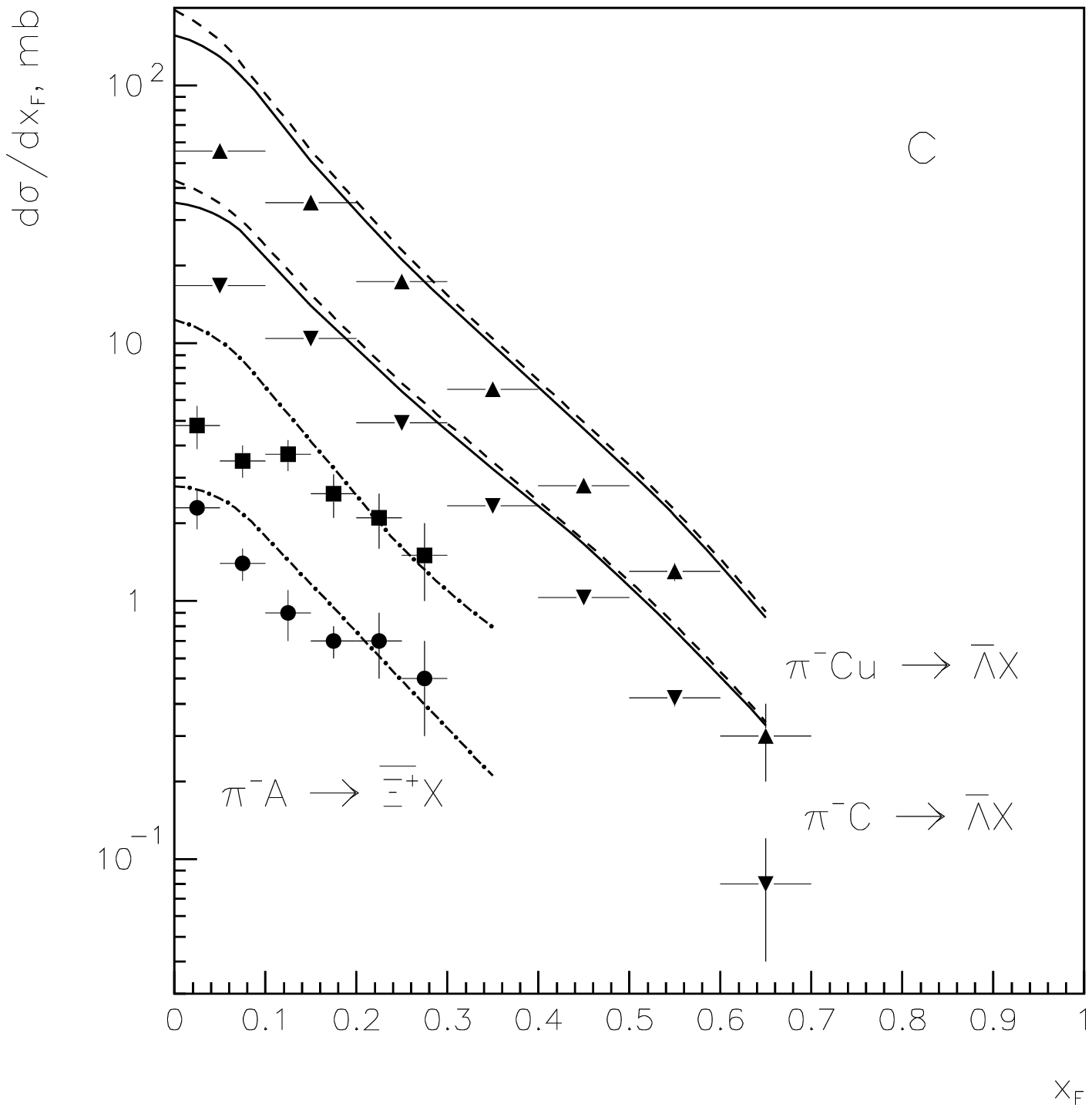}
\vskip -1cm
\caption{Feynman-$x$ distributions of secondary $\Lambda$ (a), $\Xi^-$
(b), $\bar{\Lambda}$ and $\overline{\Xi^+}$ (c) produced in $\pi^-C$
and $\pi^-Cu$ interactions at 345 GeV/c. Solid and dashed curves show
the QGSM prediction for secondary hyperon spectra with and without SJ
contribution. The QGSM predictions for antihyperon production in (c)
are shown by dash-dotted curves (b)
.}
\end{figure}

The yields of $\bar{\Lambda}$ and $\overline{\Xi^+}$ \cite{WA89}, which
do not depend on SJ contribution, are shown in Fig.~7c. These data are
described by QGSM on the reasonable level.

The data presented in \cite{WA89} allow one to calculate the asymmetries
of secondary $\Lambda / \bar{\Lambda}$ production defined as
\begin{equation}
A_{\Lambda} = \frac{N_{\Lambda} - N_{\bar{\Lambda}}}
{N_{\Lambda} + N_{\bar{\Lambda}}}
\end{equation}
They are presented in Fig.~8a for the cases of $\pi^-Cu$ (points)
and $\pi^-C$ (squares) interactions. The curves show the QGSM
calculations for the cases of copper (dotted curve), carbon (dashed
curve) and nucleon (solid curve) targets. We predict some $A$-dependence
of the asymmetry for beam fragmentation region. The agreement with the
data is good in the central region, but 
the asymmetry is underestimated in the forward region.

In Fig.~8b we present the data of \cite{E769} for the same asymmetry
Eq.~(13) obtained for $\pi^-$ interactions with multifoil target with
different atomic weights, see \cite{E769}. Here the QGSM predictions
even for $\pi^-p$ interactions (solid curve), i.e. neglecting the
$A$-dependence, overestimate the data at $x_F > 0.1$. In the central
region, $\vert x_F \vert \leq 0.1$, our calculations are in agreement
with the data of both \cite{WA89} and \cite{E769} as well as of
\cite{ait}. Here we predict the practical absence of $A$-dependence
(or weak dependence) for $\Lambda / \bar{\Lambda}$ asymmetries, as it
was assumed in \cite{ACKS}.

\begin{figure}[htb]
\centering
\vskip -2.cm
\includegraphics[width=.49\hsize]{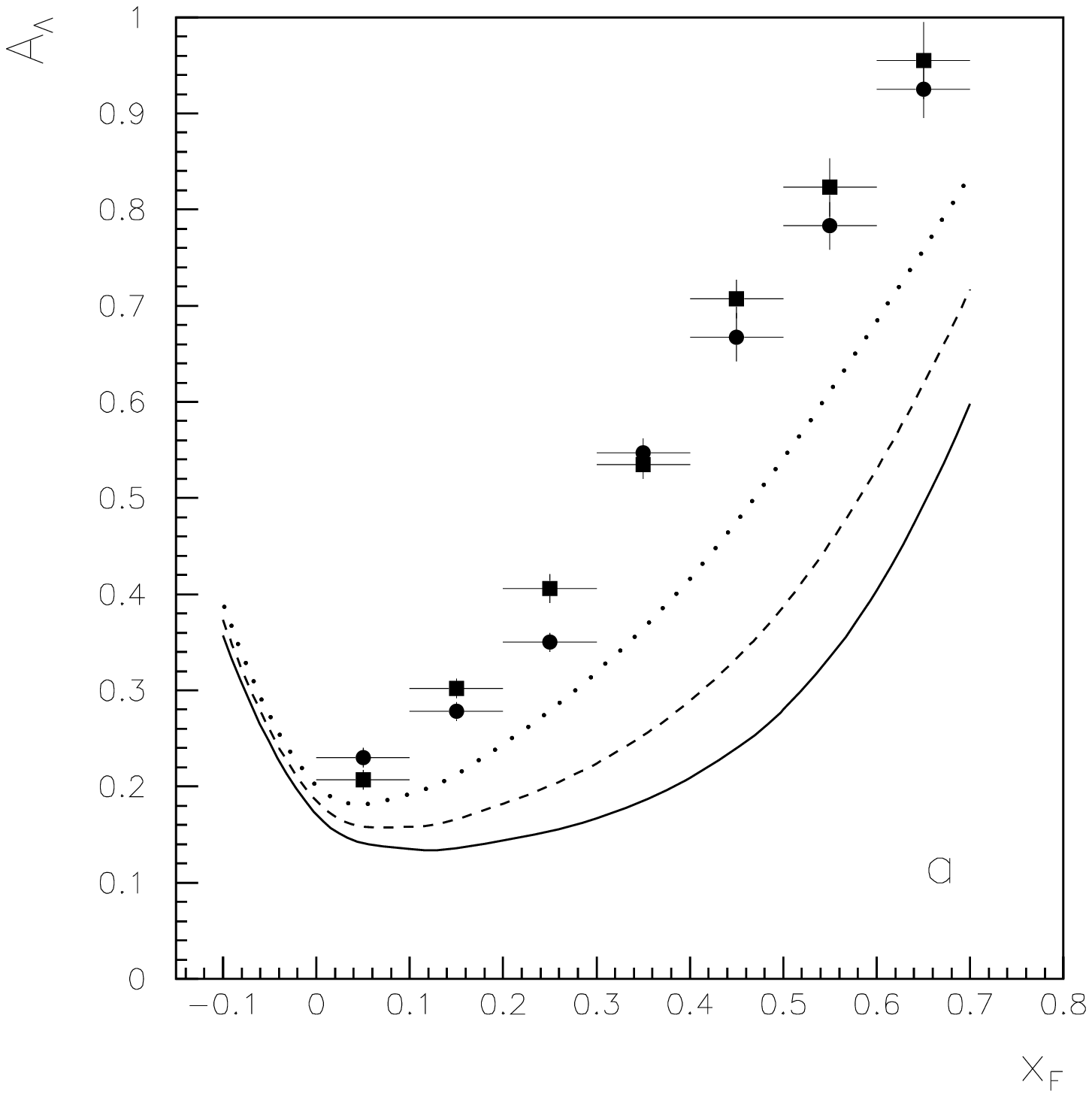}
\includegraphics[width=.49\hsize]{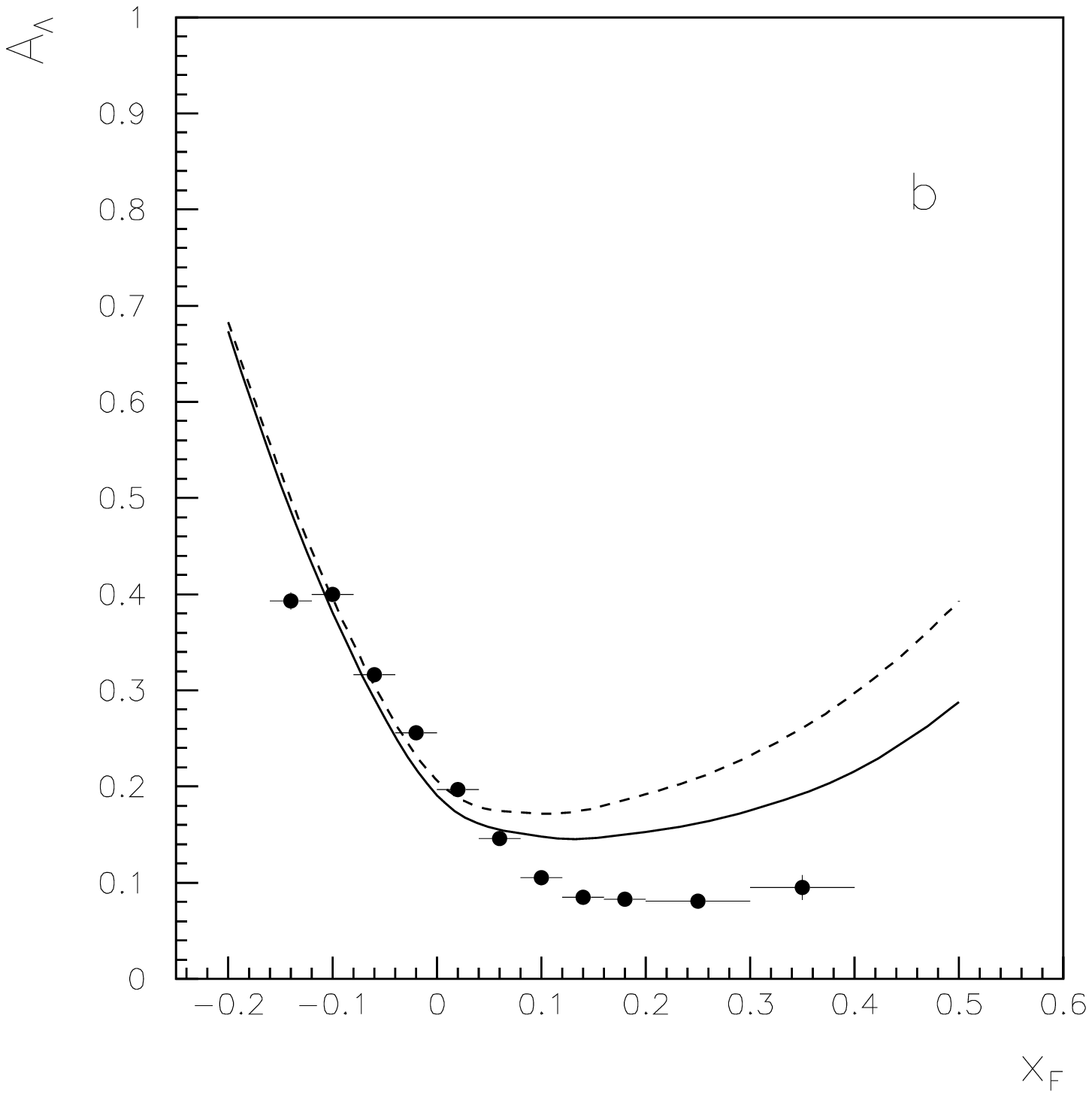}
\vskip -1cm
\caption{The asymmetries of secondary $\Lambda/\bar{\Lambda}$
production in $\pi^-C$ (squares) and $\pi^-Cu$ (points) interactions
at 345 GeV/c (a). The same asymmetries for $\pi^-A$ collisions at
250 GeV/c (b). Solid, dashed and dotted curves show the QGSM predictions
for nucleon, carbon and copper targets, respectively.
}
\end{figure}

The comparison of the data shown in Figs.~7a and 7c allows us to obtain
the direct results for SJ contribution to hyperon production cross
section. 
$\Lambda$ has the valence quark content $uds$, so the
fast incident $\pi^-$ $(\bar{u}d)$ should fragment into secondary
$\Lambda$ and $\bar{\Lambda}$ with equal probabilities, i.e. the
$\pi^- \to \Lambda,$ resp. $\pi^- \to \bar{\Lambda}$ fragmentation is flavour
symmetrical, contrary, say, to the case of $\pi^- \to p$ resp. $\pi^- \to
 \bar{p}$
fragmentation.

%
The contributions of the processes of Fig.~3a and 3b are negligible
at $x_F > 0.1$ and the difference in the spectra of secondary $\Lambda$
and $\bar{\Lambda}$ determines the SJ contribution of the process shown
in Fig.~3c. 

This difference is obtained to be rather large in
\cite{WA89} but very small in \cite{Mik}. To show the disagreement
between the data of \cite{WA89} and \cite{Mik} we present in the table
the values of the parameter $n$ for the parametrization
\begin{equation}
d\sigma/dx_F = C (1-x_F)^n \\ ,
\end{equation}
which were obtained in \cite{WA89} and \cite{Mik} for secondary $\Lambda$
and $\bar{\Lambda}$ production.

\newpage

\begin{center}
{\bf Table 1}
\end{center}
\vspace{15pt}
The values of $n$ parameter in Eq.~(14) obtained in \cite{Mik} and
\cite{WA89} for $\Lambda$ and $\bar{\Lambda}$ production in high
energy $\pi^-p$ and $\pi^-A$ collisions.
\begin{center}
\vskip 12pt
\begin{tabular}{|c|c|}\hline
 Reaction & n  \\   \hline
$\pi^-p \to \Lambda$  \cite{Mik} & $2.0 \pm 0.1$  \\
$\pi^-C \to \Lambda$  \cite{WA89} & $2.12 \pm 0.02$   \\
$\pi^-Cu \to \Lambda$ \cite{WA89} & $2.71 \pm 0.02$   \\
$\pi^-p \to \bar{\Lambda}$  \cite{Mik} & $2.0 \pm 0.1$  \\
$\pi^-C \to \bar{\Lambda}$  \cite{WA89} & $5.23 \pm 0.04$  \\
$\pi^-Cu \to \bar{\Lambda}$  \cite{WA89} & $5.53 \pm 0.04$  \\

\hline
\end{tabular}
\end{center}

\vskip 0.5cm

The values of $n$ for the secondary $\Lambda$ production obtained both
in \cite{WA89} and \cite{Mik} on nucleon and on nuclear targets are in
agreement with a natural weak $A$-dependence. The value of $n$
slightly increases with $A$ that demonstrates a well-known effect of
nuclear absorption \cite{KTMS,Sh2,CK,ASS}. The values of $n$ for
$\bar{\Lambda}$ production obtained in \cite{Mik} and \cite{WA89} are
absolutely different. The data of \cite{Mik} show the absense or very
small contribution of SJ diffusion in the case of $\Lambda$ and
$\bar{\Lambda}$ production, in contradiction with another results, see
for example \cite{ait}.

It is possible to extract the SJ contribution from the experimental
data of \cite{WA89} only. 
The SJ contributions to the spectrum of secondary
$\Lambda$ in $\pi^-Cu$ and $\pi^-C$ collisions, obtained by such a way
are presented in Fig.~9.

\begin{figure}[htb]
\centering
\vskip -1cm
\includegraphics[width=.55\hsize]{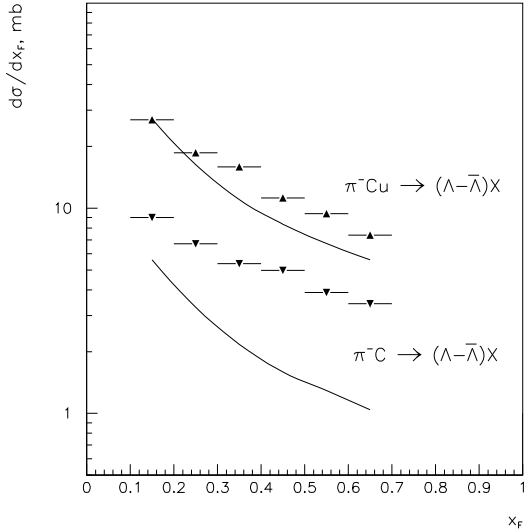}
\vskip -1cm
\caption{The extracted SJ contributions to the spectra of $\Lambda$
in $\pi^-A$ collisions at 345 GeV/c and their description by QGSM.
}
\end{figure}

The $x_F$-distributions of the $\Lambda$ produced from copper
target are in reasonable agreement with QGSM calculations, however, 
in the case of carbon target we obtain the disagreement coming
mainly from unsatisfactory
description in Fig.~7a and also from some
overestimation of $\bar{\Lambda}$ production in Fig.~7c. 
A reminder, the data \cite{Mik} would lead to very small SJ
contribution in this region way below the model.

There exists only a few data on secondary production in 
nucleon-nucleon and
nucleon (deuteron) -nucleus collisions at RHIC energies. In Fig.~10a
we present the rapidity (in c.m.) distribution of the ratio
$\bar{p}/p$ in $pp$ interactions at $\sqrt{s}$ = 200 GeV \cite{BRA}.
The QGSM calculation with the same SJ contribution (solid curve) is in
reasonable agreement with the data, and the same calculations without
SJ contributions (dashed curve) overestimate the discussed ratios.

In Fig.~10b the dependence of $\bar{p}/p$ ratios at
$\vert y_{c.m.} \vert =0$ in
$dAu$ collisions is shown as a function of ``centrality" 
at $\sqrt{s}$ = 200 GeV \cite{PHOB}. The experimental
data are shown here by open squares and the QGSM predictions with SJ
contribution by the solid curve which is very close to the constant.
The calculations without SJ contribution (dashed curve) again lead 
to high values of $\bar{p}/p$ ratios. The close points in Fig.~9b
present the predictions of the DPMJET-III model \cite{DPMJET} and they
are in agreement with the data as well as with QGSM calculations.
Dash-dotted curves in Fig.~10a, b show the QGSM predictions for
$\bar{\Lambda} / \Lambda$ ratios.

\begin{figure}[htb]
\centering
\vskip -1.cm
\includegraphics[width=.49\hsize]{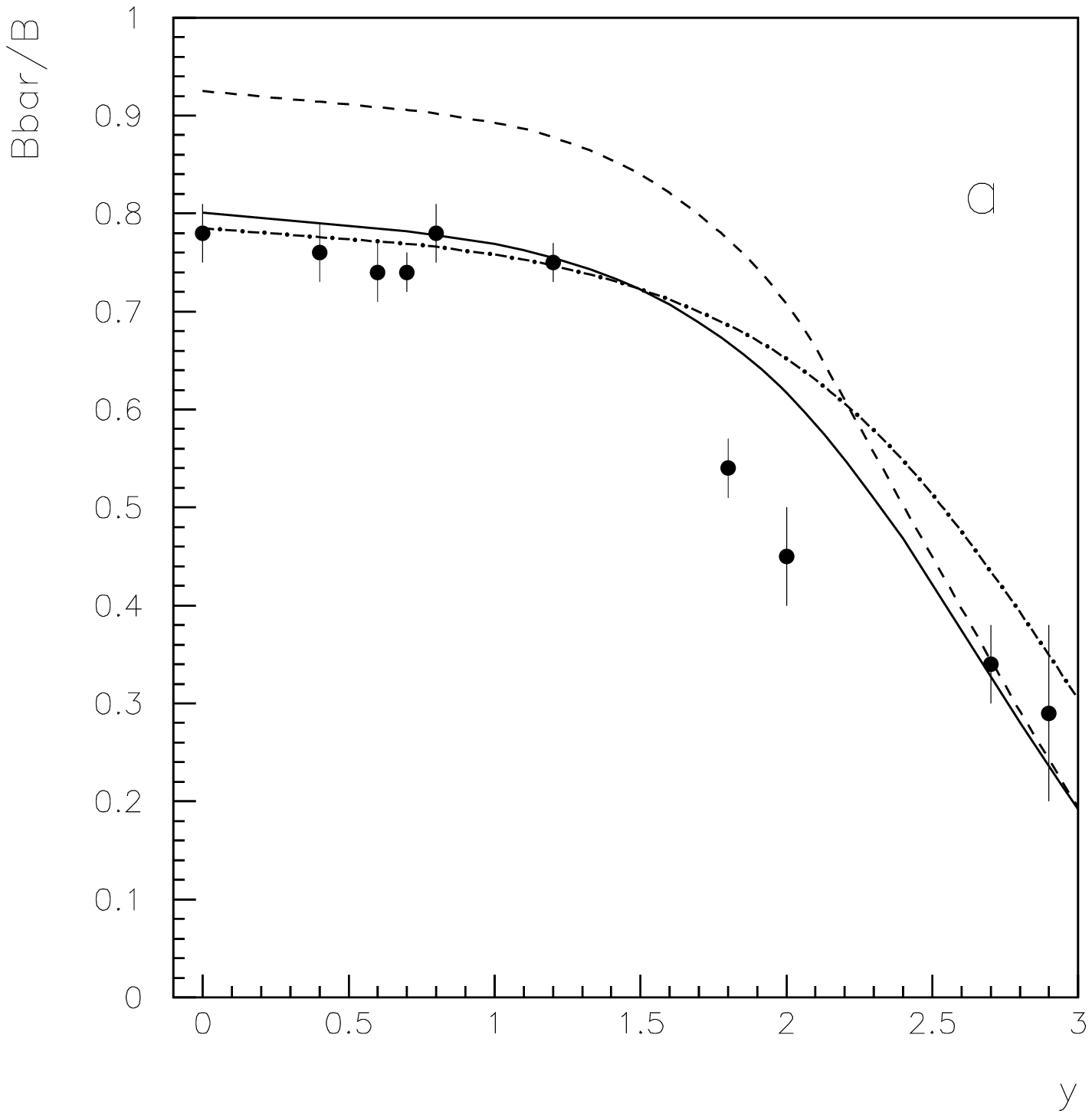}
\includegraphics[width=.49\hsize]{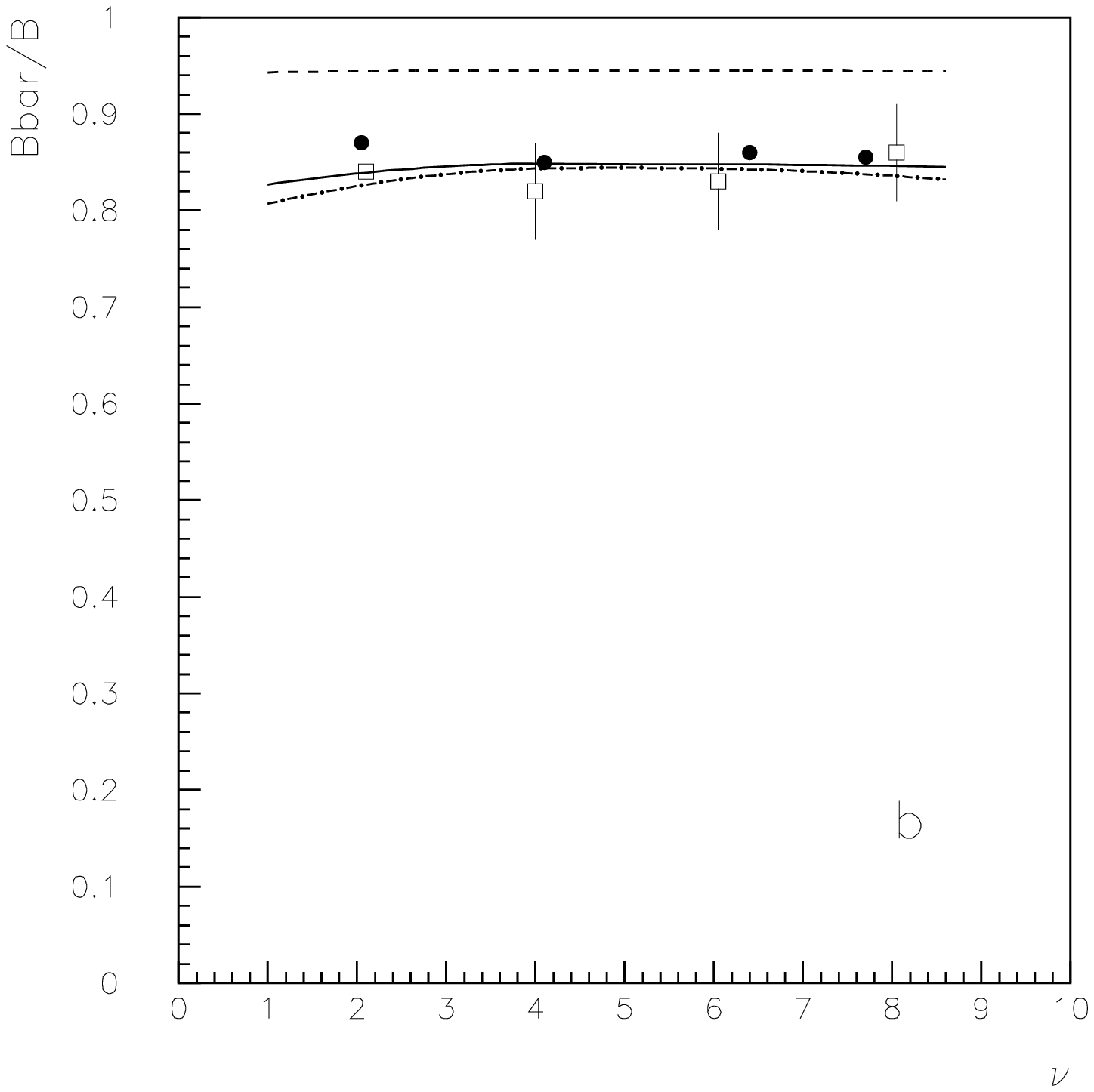}
\vskip -1cm
\caption{The rapidity dependence of $\bar{p}/p$ ratios for $pp$
collisions at $\sqrt{s}$ = 200 GeV. Solid and dashed curves show
the QGSM description with and without SJ contribution and the dash-dotted
curve shows the QGSM predictions for $\bar{\Lambda} / \Lambda$ ratio
(a). The $\bar{p}/p$ ratio as a function of ``centrality" for $dAu$
collisions at $\sqrt{s}$ = 200 GeV (open squares) together with the
QGSM with SJ (solid curve) and without SJ (dashed curve) and with the
DPMJET-III model (closed points) predictions. The QGSM predictions
for $\bar{\Lambda} / \Lambda$ ratio are shown by dash-dotted curves.}
\end{figure}

The predictions of several other models \cite{HIJING,RQMD,AMPT} are
in some disagreement with the data of \cite{PHOB} (see Fig.~4 in
\cite{PHOB}). The extrapolation of the predictions of these models to
$\nu = 1$ give the values of $\bar{p}/p$ in $pp$ interactions larger
than 0.9 that contradicts the data presented in Fig.~9a.

\vskip 0.4 truecm
\noindent{\bf 5. CONCLUSIONS}
\vskip 0.4 truecm

We presented the role of string junction diffusion for the baryon charge
transfer over large rapidity distances mainly for the case of collisions
with nuclear targets. Without this contribution shown in Fig.~3c the
data for baryon/antibarion yields and asymmetries are in disagreement
with the data. The discussed string junction effects has $A$-dependences
which are in general in agreement with the QGSM predictions (see, for
example, Fig.~6).

It is necessary to note that the existing experimental data are not
enough for determination of the SJ parameters with the needed
accuracy. Some data are in disagreement with other ones, for example,
the behaviour of $\bar{\Lambda}$ spectra at $x_F > 0$ obtained by
\cite{Mik} and \cite{WA89}, see Table, and the $\Lambda/\bar{\Lambda}$
asymmetries in \cite{WA89} and \cite{E769} which are presented in
Figs.~8a and 8b.

We are grateful to J. Ranft, R. Engel, G. H. Arakelyan, A. B. Kaidalov, L. N. Lipatov,
O. I. Piskounova and A. A. Rostovtsev for useful discussions.
This paper was supported by DFG grant GZ: 436 RUS 113/771/1-2 and, in
part, by grants RSGSS-1124.2003.2 and PDD (CP) PST.CLG980287.




\end{document}